\documentclass[prd,aps,twocolumn,superscriptaddress,groupedaddress,notitlepage,nofootinbib]{revtex4-2}
\usepackage{amssymb,amsthm,amsmath}
\usepackage{textcomp}
\usepackage{color}      
\usepackage{slashed}    
\usepackage{verbatim}
\usepackage{hyperref}
\usepackage[normalem]{ulem} 

\usepackage{soul}

\usepackage{booktabs}
\usepackage{rotating}   
\usepackage{multirow}   
\usepackage{simpler-wick}

\usepackage{subcaption}
\usepackage{epsfig}
\begin{document}
	
	\raggedbottom
	
	\title{
Potential of discovering $\Xi_{cc}^+$ in $\Lambda_b $ decays }

\author{   Chao-Qiang Geng}	
\affiliation{School of Fundamental Physics and Mathematical Sciences, Hangzhou Institute for Advanced Study, UCAS, Hangzhou 310024, China}
\author{   Xiang-Nan Jin } 
\affiliation{School of Fundamental Physics and Mathematical Sciences, Hangzhou Institute for Advanced Study, UCAS, Hangzhou 310024, China}
\author{   Chia-Wei Liu  }
\affiliation{School of Fundamental Physics and Mathematical Sciences, Hangzhou Institute for Advanced Study, UCAS, Hangzhou 310024, China}
\author{   Xin-Yi Liu  }
\affiliation{School of Fundamental Physics and Mathematical Sciences, Hangzhou Institute for Advanced Study, UCAS, Hangzhou 310024, China}
\author{   Xiao Yu }

\affiliation{School of Fundamental Physics and Mathematical Sciences, Hangzhou Institute for Advanced Study, UCAS, Hangzhou 310024, China}

\date{\today}

\begin{abstract}

 We propose to search for $\Xi_{cc}^+$ in the decay $\Lambda_b \to \Xi_{cc}^+ D^-$, which serves as a tagged and reconstructible source of $\Xi_{cc}^+$, providing an experimentally clean environment for its discovery. A possible fully charged final state is $[(pK^- K^- \pi^+)_{\Xi^0_c} \pi^ + ]_{\Xi_{cc}^+} \,(K^+ \pi^- \pi^-)_{D^-}$, where the subscripts indicate the parent hadron in the decay chain. We estimate the branching fraction to be ${\cal B}(\Lambda_b \to \Xi_{cc}^+ D^-) = ( 5.1 \pm 3.9) \times 10^{-4}$, while the fully charged final state corresponds to an effective branching fraction of order $10^{-8}$. These results indicate that LHCb has the potential to discover the $\Xi_{cc}^+$. The prospects for observing $\Omega_{cc}^+$ are also briefly discussed.
\end{abstract}
\maketitle

\section{
Introduction
}

In 2017, the LHCb collaboration reported the first   observation of the doubly charmed baryon 
$\Xi_{cc}^{++}$~\cite{LHCb:2017iph,Yu:2017zst}, opening a new frontier in heavy-flavor physics. 
Its mass was measured to be 3621.55 MeV~\cite{LHCb:2019epo}, with a lifetime of 0.256 ps~\cite{LHCb:2018zpl}. Much earlier, the SELEX experiment claimed an evidence for   $\Xi_{cc}^{+}$, observed in the decay channels 
$\Lambda_{c}^{+} K^{-} \pi^{+}$ 
and 
$pD^{+}K^{-}$, 
with a reported mass of 3518.7 MeV and an extremely short lifetime~\cite{SELEX:2002wqn}. However,   BaBar, Belle, and LHCb were unable to confirm the observation~\cite{BaBar:2006bab,Belle:2013htj,LHCb:2013hvt}.  
The long-awaited evidence for $\Xi_{cc}^+$ and $\Omega_{cc}^+$ remains absent.

The existence of two heavy quarks within a hadron provides a unique playground for studying heavy-quark dynamics. Numerous approaches have been devoted to exploring the properties of doubly charmed baryons~\cite{Detmold:2015aaa,Wang:2017mqp,Cheng:2018mwu,Cheng:2020wmk,Tong:2021raz,Zeng:2022egh,Liu:2023lsg,Liang:2023scp,Geng:2022uyy,Geng:2022xfz}. In particular, theoretical investigations of $\Xi_{cc}^{+}$ are of significant importance. 
The differences in the theoretical properties of $\Xi_{cc}^{+}$ and $\Xi_{cc}^{++}$, such as their lifetimes~\cite{Kiselev:1998sy,Likhoded:1999yv,Chang:2007xa,Dulibic:2023jeu}, can provide clear benchmarks for the experimental discrimination.  
Thus, 
it is crucial to explore new production mechanisms for doubly charmed baryons that can be realized in current or forthcoming experiments~\cite{Xing:2023kjk}. 
The decays of heavy hadrons such as the $\Lambda_{b}^{0}$ baryon   provides natural environments for producing final states containing three charm quarks. 
It offers a tagged and kinematically constrained production mechanism for $\Xi_{cc}^+$, enabling a clean and feasible search at LHCb.
In these processes, short-distance contributions from $W$-exchange diagrams are typically suppressed, whereas long-distance effects, especially final-state interactions (FSI), are expected to play a dominant role~\cite{Fajfer:2003ag,Jia:2024pyb,Geng:2024uxp,Hu:2024uia}.  
Motivated by these considerations, we investigate the decay channel 
$\Lambda_{b} \to \Xi_{cc}^{+} D^{-}$ 
as a   production source of $\Xi_{cc}^{+}$. 
We focus on the rescattering   from 
$\Lambda_{b}^{0} \to \Lambda_{c}^{+} \pi^{-}$, 
which is color-enhanced and well measured experimentally~\cite{LHCb:2014ofc}, depicted in FIG.~\ref{STU}. 

The structure of this paper is as follows. In Sec.~\ref{2}, we introduce the theoretical framework and formalism of the decay amplitudes. In Sec.~\ref{3}, we present the branching ratios of different decay channels as well as the parameters $\alpha$, $\beta$, and $\gamma$, followed by a detailed discussion. We make conclusion in Sec.~\ref{4}.

\section{
Formalism 
}
\label{2}
The amplitude
of 
$\Lambda_b ^0 \to \Xi_{cc}^+  D ^-$ 
reads  
\begin{equation}
	{\cal M} =   i \overline{u} _{cc} ( A + B \gamma _5 )  u _ b \,. 
\end{equation}
At the quark level, this process is induced by 
the  $W$-exchange diagram with a pair of charmed quarks
produced from the vacuum. Thus, the short-distance 
contributions are expected to be tiny, as 
the relevant gluon propagator is suppressed by $1/( 4m_c^2) $. 
On the other hand, the released energy 
$m_{\Lambda_b}
- m_{\Xi_{cc}} - m_D < m _\pi
$ 
is very small, and the long-distance FSI is expected to dominate.

\begin{figure}
\includegraphics[width=0.23 \textwidth]{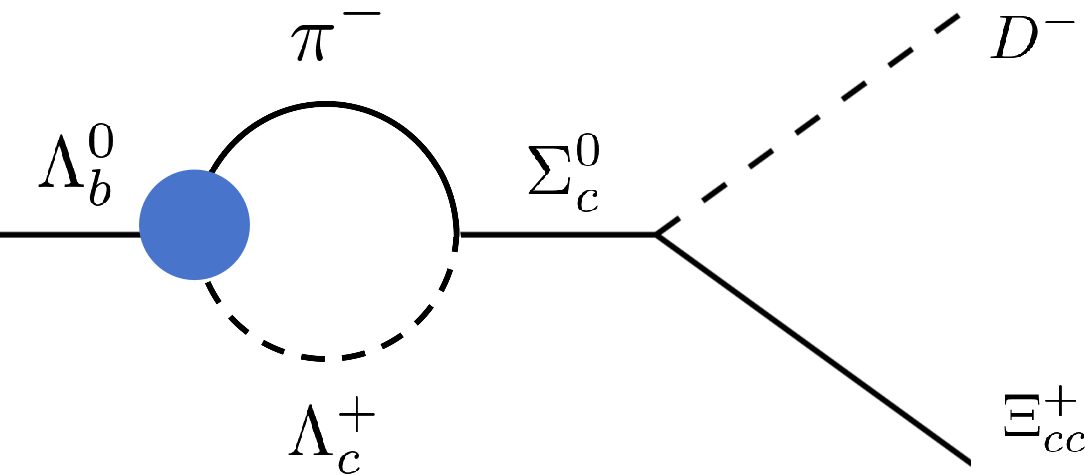}
\includegraphics[width=0.23 \textwidth]{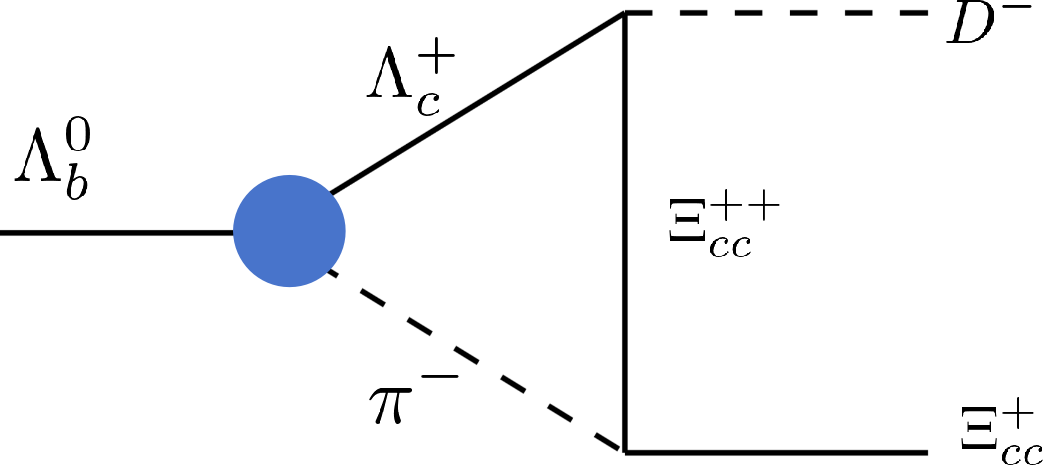}
\includegraphics[width=0.23 \textwidth]{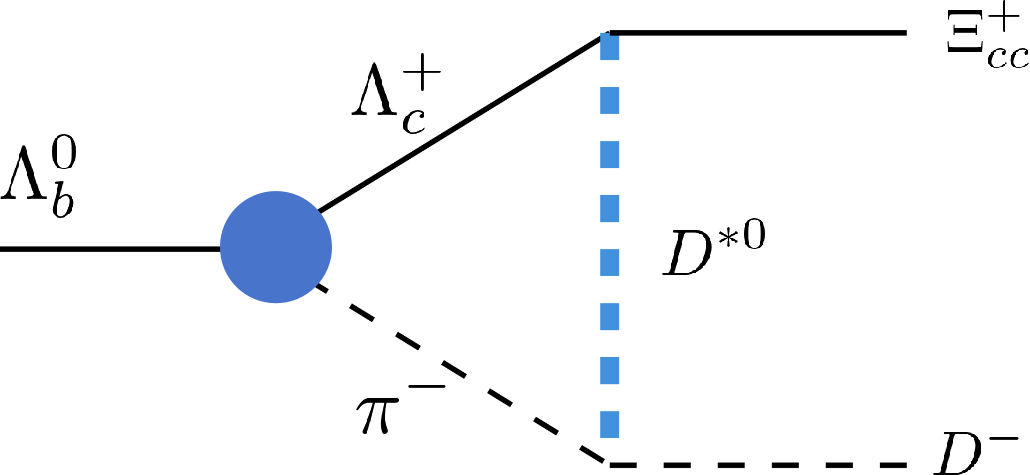}
\caption{
The FSI diagrams for the $s$-, $t$-, and $u$-channel rescattering. 
	}
\label{STU}
\end{figure}

In this work, we consider 
FSI from 
$\Lambda_b ^0 \to \Lambda_c ^+ \pi ^ - $. We choose this channel because 
it has already been measured experimentally and is color-enhanced.
Other possible intermediate states, such as 
$\Lambda_b \to \Lambda_c^+ \rho ^-, \Lambda_c^+ a(1260)^-$, have not yet been measured and may be considered after such measurements become available.
There are 
three possible FSI diagrams, depicted in FIG.~\ref{STU}. The blob represents the insertion of the effective Lagrangian that induces  $\Lambda_b  \to \Lambda_c^+ \pi ^- $, given by
\begin{equation}
	{\cal L} _{\text{eff}} = 
	- \frac{
		G_F
	}{\sqrt{2} }
	V_{cb} V_{ud} ^ * 
	\left[ 
	c_1  
(\overline{c} b)_L ( \overline{d} u) _L
	+
	c_2 
(\overline{d} b)_L ( \overline{c} u) _L
	\right]  \,,
\end{equation}
where 
$ (\overline{q}_4 q_3)_L ( \overline{q}_2 q_1) _L
=
(\overline{q}_4\gamma_\mu ( 1-\gamma_5)  q_3)  ( \overline{q}_2 \gamma^\mu ( 1-\gamma_5)  q_1) $, and  $c_1$ and $c_2$ are 
the Wilson coefficients. At the hadron level, 
it is decomposed as  the $S$ and $P$ waves
${\cal L}_{\text{eff}}
= - i  \pi ^ + \overline{\Lambda}_c  \left(
a + b \gamma _5 
\right) \Lambda_b \,. 
$
To study the running of $a$ and $b$ with respect to momenta, we adopt the naive factorization for 
$\Lambda_ b ^ 0 \to \Lambda_c ^ + \pi ^-$, which reads 
\begin{eqnarray}
	\overline{u}_c 
	(a + b \gamma_5 ) 
	u _b 
	&=& \xi 
	q ^ \mu \langle \Lambda_c ^+ | \overline{c} \gamma_ \mu ( 1 - \gamma_5) b | \Lambda_b ^0 \rangle   \,,\nonumber\\
	\xi 
	&=&  f_ \pi \frac{G_F}{
		\sqrt{2}
	}
	V_ {cb}  V_{ud} ^* 
	\left(
	c_1 + \frac{c_2}{N_c}
	\right)  \,,
\end{eqnarray}
with $N_c$ the color number. By expanding the baryon matrix by  form factors defined in 
Ref.~\cite{Detmold:2015aaa}, we obtain 
$ a = 
\xi    f_0  m_ - 
\,$ and~ $ 
b  =
\xi  g_0 
m_+ 
\,,
$
with $
m_\pm  
=    m_{\Lambda_b}  \pm  m _{\Lambda_c}   \,.
$
The running of the form factor is governed by 
a monopole behavior 
\begin{equation}\label{latticeform}
\begin{aligned}
	f (q^2) &= 
	\frac{ ( m_ {\text{pole}} ^f)  ^2  }{
		( m_ {\text{pole}} ^f)  ^2    - q ^2 } 
	\left(
	a _0 ^f + a_1 ^f 
z_f(q^2)
	\right)   \,,    \\
	z_f(q^2) &=  	\left(
	\frac{
		\sqrt{
			( m_ {\text{pole}} ^f) ^2 
			-q ^2 }
		- \sqrt{
			( m_ {\text{pole}} ^f) ^2 
			- m_- ^2 
		}
	}{
		\sqrt{
			( m_ {\text{pole}} ^f) ^2 
			-q ^2 }
		+ 	\sqrt{
			( m_ {\text{pole}} ^f) ^2 
			- m_- ^2 
	}}
	\right)\,,     
\end{aligned}
\end{equation}
where 
$ f  = f_0   , g_0$, 
and $m_{\text{pole}}^f  $ 
is  given by the low-lying $B_c$ meson with the correct parity
$
( m_{\text{pole}}^{f_0} ,
m_{\text{pole}}^{g_0}) 
= (6.725 , 6.275 )~\text{GeV} \,,
$ 
while 
$( a_0^{f_0}
, a_1^{f_0},
a_0^{g_0}
, a_1^{g_0}
) = ( 0.744, -4.648,0.740, -4.366)$
are fitted from the lattice~\cite{Detmold:2015aaa}. 
Since  $|z_f(q^2)| \leq  0.1 $ 
at $ 0< q^2 < m_-^2$, 
$a_0^f$ and $a_1^f$ are known to be 
  the leading order~(LO) and the next-to-the leading order~(NLO) correction in form factors.

After extracting the common factors, the 
$S$-wave 
amplitude  is   decomposed as  
\begin{equation}\label{12}
\begin{aligned}
	A =&  
	\frac{\xi m_-}{16 \pi ^2 }
	\Big[ 
		g_{\Lambda_c ^+ \Sigma _{c}^{0}}^{\pi^-}  
	g_{\Sigma _{c}^{0} \Xi_{cc}^{+}}^{D^+} 
	{\cal S}^{A} + 
	g_{\Lambda_c ^+ \Xi_{cc}^{++}}^{D^+} 
	g_{\Xi_{cc}^{++} \Xi_{cc}^{+}}^{\pi^-} 
	{\cal T}^{A}\\
	&  \qquad \qquad +
	g_u\left(  
	F_1 {\cal U}_1^{A}
	+ F_2 {\cal U}_2^{A}
	\right) 
	\Big],  
\end{aligned}
\end{equation}
Here, ${\cal S}^{A}$  denotes the $s$-channel loop integral according to FIG.~\ref{STU}, while ${\cal T}^{A}$ and ${\cal U}^{A}$ correspond to the $t$- and $u$-channels, respectively.
They  depend only on the
hadron masses and 
 form factors. 
In the NLO corrections, $z(q^2)$ behaves like a propagator with a square-root singularity.
The $P$-wave amplitude is defined in a similar manner by replacing $m_-$ with $m_+$, and the corresponding loop integrals are denoted as ${\cal S}^B$, ${\cal T}^B$, and ${\cal U}_{1,2}^B$.

The hadronic couplings are defined by the effective Lagrangian  
\begin{eqnarray}
    {\cal L}_{\mathrm{int}} 
	&= &
-i 
	\sum g_{{\bf B}_1 {\bf B}_2}^{P}\,
P \overline{{\bf B}}_2 \gamma_5 {\bf B}_1 
-i 
 g_u 
	(\overline{D}^{*0})^\mu 
	\pi^- 
	\partial_\mu D^+    \\
    & + &
	(D^{*0})^\mu \left[
	F_1
	\overline{\Xi_{cc}^+}  
	\gamma_\mu 
	\Lambda_c^+ 
+
\frac{F_2}{m_c+ m_{cc} }
\partial^\nu 
\left( \overline{\Xi_{cc}^+}  
\sigma_{\mu\nu }
\Lambda_c^+ 
\right)
\right]. \nonumber 
\end{eqnarray}
 The physical quantities are 
 calculated as  the bilinear 
 form of $A$ and $B$, given by 
 \begin{eqnarray}
 	\Gamma& = &\frac{|\vec{p}\, |}{8\pi  }
 	\frac{\left(m_{\Lambda_b} + m _{\Xi_{cc}}\right) ^2  - m _D^2 }{m_{\Lambda_b}^2 }\left(
 	\left|
 	A
 	\right|^2 + \kappa ^2 \left|
 	B
 	\right|^2 
 	\right)\,, \nonumber\\
 	\alpha
 	&=&\frac{
 		-2 \kappa \text{Re}\left(
 		A^*   B
 		\right)
 	}{
 		\left|
 		A
 		\right|^2 + \kappa^2 \left|
 		B
 		\right|^2 
 	}\,, 
 \end{eqnarray}
 where 
 $\kappa=  
 \sqrt{(E- m_{\Xi_{cc}^+} )/ (E+m_{\Xi_{cc}^+} )}
 $ and  $E$ the energy  of $\Xi_{cc}^+$ at the rest frame of $\Lambda_b$.

\section{
Numerical results and discussions 
}
\label{3}

For the hadron couplings, we adopt the Goldberg-Triemann relation 
\begin{eqnarray}
g_{\Lambda_c ^+ \Sigma _c ^0 } ^{\pi ^- } 
&=& \frac{G^ {\pi^- }_{\Lambda_c ^+\Sigma_c ^0 } }{
f_\pi 
}
\left(
m_{\Sigma_c} + 
m_{\Lambda _c}
\right)\,, \nonumber\\
g_{\Xi_{cc} ^{++} \Xi_{cc} ^{+} } ^{\pi ^- } 
&=&  \frac{G^{\pi^-} _{\Xi_{cc} ^{++} \Xi_{cc} ^{+} }}{
	f_\pi 
}
2 m_{\Xi_{cc}} 
\end{eqnarray}
where 
$G^{\pi^-}_{{\bf B}_1{\bf B}_2}$
is the leading axial form factor. 
They are  found to be 
	$ 
( G_{\Lambda_c ^+ \Sigma _c ^0 } ^{\pi ^- }  , 
G_{\Xi_{cc} ^{++} \Xi_{cc} ^{+} } ^{\pi ^- } ) = ( 
	0.53 , -0.259) $ in the MIT bag model~\cite{Cheng:2018hwl,Liu:2022igi}. 
Hence, 
$( g_{\Lambda_c ^+ \Sigma _c ^0 } ^{\pi ^- } ,
g_{\Xi_{cc} ^{++} \Xi_{cc} ^{+} } ^{\pi ^- } )
= ( 19.3, -14.4 )$. 
On the other hand, 
the light-cone QCD sum rule~\cite{Hu:2025ajx}
gives 
$g_{\Lambda_ {c} ^{+} \Xi_{cc} ^{++} } ^{D ^+} 
= 
-2.72 \pm 0.92 \,, 
$ 
while 
$g_{\Sigma _ {c} ^ 0  \Xi_{cc} ^{+} } ^{D ^+  } $ 
is missing in the literature. 
The couplings are expected to be proportional to the leading form factors, and hence we use the approximation of 
$
g_{\Sigma _ {c} ^ 0  \Xi_{cc} ^{+} } ^{D ^+  }
/
g_{\Lambda_ {c} ^{+} \Xi_{cc} ^{++} } ^{D ^+} 
\approx 
	G_{\Sigma _ {c} ^ 0  \Xi_{cc} ^{+} } ^{D ^+  }
/
	G_{\Lambda_ {c} ^{+} \Xi_{cc} ^{++} } ^{D ^+} 
	\approx   5\sqrt{2/3 }
	$\,. The value of $ 
	5\sqrt{2/3 }$ 
	is used according to the quark model~\cite{Cheng:2020wmk}.  The 
    experimental data 
    of $D^* \to D \pi $
    gives 
    $g_u =  16.8$~\cite{ParticleDataGroup:2024cfk}.  
In the framework of the heavy quark spin symmetry,
the $D$ and $D^*$ mesons are related, and
the form factors $F_1$ and $F_2$
can be determined from the input of $g^{D^+}_ {\Lambda_c^+ \Xi_ {cc}^{++}}$, found to be 
\begin{equation}\label{HQETrelation}
	g^{D^0}_{\Lambda_c^+\Xi_{cc}^+} : F_1 : F_2 = 1 : 0.40 : 0.46.
\end{equation}
A detailed derivation is collected in Appendix \ref{HQET_app}. 


The Wilson coefficients depend on the energy scale, which are not respected in the naive factorization.
We fit  it by the experimental data of 
${\cal B}(\Lambda_b \to \Lambda_c^+ \pi^-)
= (4.9\pm0.4) \times 10 ^{-3} 
$~\cite{ParticleDataGroup:2024cfk}, which corresponds to $(c_1 + c_2/N_c ) 
=0.91 \pm 0.04 . $ 
The  numerical values
of the loop integrals 
 are collected in TABLE~\ref{tab:loops}.
We note that Im$({\cal S}) $ 
can only be  induced by the on-shell of $\Lambda_c^+ \pi^-$. Hence, it only depends on $f(q^2) $ at $q^2 = m_\pi ^2 $ and  we have the relation 
\begin{eqnarray}\label{Im_Loop}
\frac{	
\operatorname{Im}\left( 
{\cal S} ^{A} _{\mathrm{NLO}}
\right) 
}{
\operatorname{Im}\left(  {\cal S} ^{A} _{\mathrm{LO}} 
\right) 
}
&=&
 \frac{
a _ 1  ^{f_0}
}{
a _ 0 ^{f_0}
}
z_{f_0} ( m_\pi ^2 ) 
= -0.42 
\,,\nonumber\\
\frac{	
	\operatorname{Im}\left( 
	{\cal S} ^{B} _{\mathrm{NLO}}
	\right) 
}{
	\operatorname{Im}\left(  {\cal S} ^{B} _{\mathrm{LO}} 
	\right) 
}
&=&
\frac{
	a _ 1  ^{g_0}
}{
	a _ 0 ^{g_0}
}
z_{g_0} ( m_\pi ^2 ) = -0.48\,. 
\end{eqnarray}
The same relations also hold for ${\cal T}^{A,B}$. For the $u$-channel, the relation is broken slightly by 
$D^{*0} \to D^- \pi ^+ $ which is highly suppressed by the small phase space.

\begin{table}[ht]
\centering
\begin{tabular}{ccrrrr}
\hline 
&
\multicolumn{1}{c}{$   {\cal S}^{A,B} $} &
\multicolumn{1}{c}{ ${\cal T}^{A,B}$ }  &
\multicolumn{1}{c}{ ${\cal U}_1^{A,B}$ }  &
\multicolumn{1}{c}{ ${\cal U}_2^{A,B}$ }  \\
\hline   
\multirow{2}{*}{$A  $} &
$-0.93-1.34i$ & $-1.36-2.11i$ & $0.11+1.96i$ & $0.41+0.39i$  \\
&
$0.63-0.74 i$ & $1.10-1.21i$ & $-0.78+1.12i$ & $-0.20+0.22i$  \\
\multirow{2}{*}{$B  $} &   
$-0.40-0.61i$ & $-0.92-0.30i$ & $1.45+3.80i$ & $-1.60-1.68i$  \\
 &
$0.14-0.31i$ & $0.60-0.15i$ & $-1.56+1.97i$ & $1.02-0.85i$  \\
\hline
\end{tabular}
\caption{Numerical values of loop integrals 
	according to FIG.~\ref{STU} and 
	Eq.~\eqref{12}. The upper rows correspond to the results with LO form factors, while the lower rows include the NLO corrections.
}
\label{tab:loops}
\end{table} 

The NLO corrections flip the sign of the real parts of the loop integrals, suggesting that one might also expect NNLO effects—by adding the $a_2^f z_f^2(q^2)$ corrections into \eqref{latticeform}—to significantly alter the results.
At present, form factors at higher orders suffer from large uncertainties due to the smallness of $z_f(q^2)$ in $0<q^2<m_-^2$. In particular, the uncertainties of $a_2^f$ exceed their central values by an order of magnitude~\cite{Detmold:2015aaa}. In the future, if form factors can be extracted with higher precision, an NNLO analysis will be highly desirable. 
On the other hand, as shown in  \eqref{Im_Loop}, the imaginary parts of the loop integrals are fixed at $q^2 = m_\pi^2\sim 0 $, a point that has already been well determined by the lattice. NNLO corrections are therefore expected to modify the imaginary parts only slightly. Since the real and imaginary parts of the loop integrals are generally expected to share the same size, a large NNLO correction is unlikely.

The branching fraction and the Lee--Yang parameter  of  
$  \Lambda_b \to \Xi_{cc} ^+
D^ - $
is  found to be 
\begin{equation}\label{numB}
	{\cal B} = (5.1 \pm 3.9)   \times 10^{-4}, ~
	\alpha    = -0.18 \pm 0.10 \,.
\end{equation}
Its branching fraction is one order smaller than that of the 
parent process
$\Lambda_b \to \Lambda_c^+ \pi^+$, as expected in the FSI expansion. 
The main uncertainties arise from 
$g^{D^+}_{\Lambda_c^+ \Xi_{cc}^{ ++} }$. 
On the other hand, to 
search for $\Omega_{cc}^+$ in Cabibbo-favored decays,   $\Xi_{b} ^0 \to  D^- \Omega_{cc} ^+ $ is an ideal option. However, the hadronic couplings involving FSI are poorly known  at the current stage, and a thorough analysis is not yet feasible. 
Here, we consider the $U$-spin symmetry, which 
relates the $d$ and $s$ quarks.  
The amplitudes then read as 
\begin{align}
\lambda 	\langle
	D^- \Xi_{cc}^ + | {\cal H}_{eff} |
	\Lambda_b\rangle 
	&= 	\langle
	D^-_s  \Omega _{cc}^ + | {\cal H}_{eff} |
	\Xi _b^0\rangle 
	= 
	\lambda A_ 1 
	  \\
\lambda 	\langle
D_s ^- \Xi  _{cc}^ + | {\cal H}_{eff} |
\Xi _b^0\rangle  	&=    
	\langle
	D^- \Omega_{cc}^ + | {\cal H}_{eff} |
	\Lambda_b\rangle 
= 
	\frac{\lambda }{2}
	\left(
	A_ 1 - A_ 0 
	\right)\,,\nonumber\\
\lambda 	\langle
D^- \Omega _{cc}^ + | {\cal H}_{eff} |
\Xi _b^0\rangle &= 
	\langle
	D^-_ s \Xi_ {cc}^ + | {\cal H}_{eff} |
	\Lambda_b\rangle 
 = 
	\frac{\lambda }{2}
	\left(
	A_ 1 +  A_ 0 
	\right)\,, \nonumber 
\end{align}
where $\lambda = V_{us}^* / V_{ud}^*$, and the subscripts of $A_{0,1}$ denote the
$U$-spin
 representations of the final states. The direct relation
between $\Lambda_b \to \Xi_{cc}^+ D ^-$ allows us to obtain $\Xi_b^0 \to \Omega_{cc}^ + D_s^-$ as a  byproduct. While the Lee--Yang parameter is  the same as those in $\Lambda_b \to \Xi_{cc}^+ D ^-$, the decay width is suppressed by $\lambda^2$, resulting in  
\begin{equation}
	{\cal B} 
	( 
	\Xi_b^0 \to \Omega_{cc}^ + D_s^-
	) = ( 2.7 \pm 2.0 ) \times 10 ^{-5}\,. 
\end{equation}
On the other hand, assuming $A_1 \gg A_0 $, 
the Lee-Yang parameters 
of other decays 
would also  be  identical to the ones in 
\eqref{numB} and 
the branching fractions are 
\begin{equation}
	 {\cal B} 
	( \Xi _b^0 \to \Omega _{cc}^+ D ^- ) 
	=
	{\cal B} 
	( \Xi _b^0 \to \Xi _{cc}^+ D ^-_s  ) 
=   1.3 \pm 1.0  \,, 
\end{equation}
in units of $10^{-4}$. The large
 branching fractions 
 indicate that they
  may be observed in future experiments.

We briefly discuss the reconstruction of the decay
$\Lambda_b \!\to\! \Xi_{cc}^+ D^-$.
The LHCb dataset is expected to contain approximately $\mathcal{O}(10^{12})$ produced $\Lambda_b$ baryons in total~\cite{CMS:2012wje,Kramer:2018rgb}.
From  
$\mathcal{B}(\Lambda_b\!\to\!\Xi_{cc}^+ D^-)= (5.1\pm3.9)\times10^{-4}$,
one anticipates a raw yield of order $10^{8}$ decays before acceptance and efficiencies.
We consider tagging $\Xi_{cc}^+$ through the two-body decay
$\mathcal{B}(\Xi_{cc}^+\!\to\!\Xi_c^0\pi^+) = (  8.12 \pm 0.55)\% $
~\cite{Liu:2022igi}. 
The bachelor $\pi^+$ and the narrow two-body kinematics provide a clean handle and improve background rejection relative to multibody tags.
On the other hand, $D^-$ is tagged by  
\(
D^-\!\to\!K^+\pi^-\pi^-
\) with $\mathcal{B}\simeq 9.4\%$~\cite{ParticleDataGroup:2024cfk},
and reconstruct $\Xi_c^0$ by  $
\Xi_c^0 \to 
f_{\Xi_c^0}$.
Here, we consider the charged and hyperon
chain 
 final states
 \begin{align}
	&\text{(C): }  
	\Xi_c^0 \to pK^-K^-\pi^+ , 
	\label{eq:xic_allcharged}\\[2pt]
	& \text{(H): }  
	\Xi_c^0 \to \Xi^- \pi^+,\ 
	\Xi^- \to \Lambda\pi^-,\ 
	\Lambda \to p\pi^- .
	\label{eq:xic_hyperon}
\end{align}
The corresponding \emph{visible} branching fraction is
\begin{eqnarray}
		\mathcal{B}_{\rm vis}(f_{\Xi_c^0})& = &
		\mathcal{B}(\Lambda_b\!\to\!\Xi_{cc}^+D^-)\,
		\mathcal{B}(\Xi_{cc}^+\!\to\!\Xi_c^0\pi^+)\,\nonumber\\
	&\times &
		\mathcal{B}(D^-\!\to\!K^+\pi^-\pi^-)\,
		\mathcal{B}(\Xi_c^0\!\to f_{\Xi_c^0})\,
	\label{eq:Bvis_2body_generic}
\end{eqnarray} 
Inserting the numerical inputs from    the data~\cite{ParticleDataGroup:2024cfk}, 
we find the following visible branching fractions:
$\text{(C) }  \mathcal{B}_{\rm vis}\!\simeq\!(1.9\pm 1.5 )\!\times\!10^{-8 },$ and 
$\text{(H) }  \mathcal{B}_{\rm vis}\!\simeq\!(3.5\pm2.8)\!\times\!10^{-8 }.
$
One obtains
$
	\text{(C)}\quad  N_{\rm sig}\ \sim\ 1.9\times10^{4}\,\varepsilon_{\rm tot},$ and 
$	\text{(H)}\quad  N_{\rm sig}\ \sim\ 3.5\times10^{4}\,\varepsilon_{\rm tot}$, respectively.
For representative efficiencies,
\(
\varepsilon_{\rm tot}=10^{-2}\Rightarrow N_{\rm sig}\sim 200\text{ and }350,
\)
for cases (C) and (H), respectively.
Given the long-lived hyperons in (H) and the all-charged topology in (C), 
the two tags are complementary: 
(C) offers compact vertices and high tracking efficiency, 
while (H) features displaced vertices aiding background suppression. 
Both profit 
mass-constrained fits to $\Xi_c^0$ and $D^-$ 
and the narrow recoil-mass peak of the two-body 
$\Xi_{cc}^+\!\to\!\Xi_c^0\pi^+$ decay, 
resulting in a cleaner signal and improved mass resolution.

On the other hand, it is possible to observe $\Omega_{cc}^+$
in $\Xi_b^0 \to \Omega_{cc}^+ D^-$
in the near future. A possible tag for $\Omega_{cc}^+$ could be
$\Omega_{cc}^+ \to \Omega_c^0 \pi^+$,
predicted to have a branching fraction of $3.96\%$~\cite{Cheng:2020wmk}.
The production of $\Xi_b^0$ is expected to be about an order of magnitude smaller than that of $\Lambda_b$~\cite{LHCb:2014ofc},
yet still sufficient given the enormous $\Lambda_b$ yield at the LHC.
To tag $\Omega_c^0$, one can exploit 
$\Omega_c^0 \to \Omega^- \pi^+$ and $\Omega_c^0 \to \Xi^- K^- \pi^+ \pi^+$.
However, the branching fractions of these sequential decays are not yet well established at the current stage.

\section{
Conclusion
}\label{4}
 
We have investigated the potential of observing the doubly charmed baryon $\Xi_{cc}^+$ through the weak decay $\Lambda_b \!\to\! \Xi_{cc}^+ D^-$. 
By incorporating final-state interactions from the color-enhanced channel $\Lambda_b \!\to\! \Lambda_c^+ \pi^-$, we have estimated 
the branching fraction to be $\mathcal{B}(\Lambda_b \!\to\! \Xi_{cc}^+ D^-) =  (5.1\pm3.9)\times10^{-4}$. 
Although the available phase space is limited, the corresponding rate remains within the measurable range at LHCb. 

Using the two-body decay $\Xi_{cc}^+\!\to\!\Xi_c^0\pi^+$ as a tag, we have found the visible branching fractions to be 
$\mathcal{B}_{\rm vis}(C)\simeq(1.9\pm1.5)\times10^{-8}$ and 
$\mathcal{B}_{\rm vis}(H)\simeq(3.5\pm2.8)\times10^{-8}$ for the charged and hyperon modes of $\Xi_c^0$, respectively. 
With ${\cal O}(10^{12})$ produced $\Lambda_b$ baryons, one expects $\mathcal{O}(10^2)$ reconstructed events after efficiencies, 
sufficient to establish evidence of $\Xi_{cc}^+$ and measure its mass with sub-MeV precision. 
The complementary tagging strategies—fully charged versus hyperon-chain reconstruction—provide distinct experimental advantages 
in tracking and background suppression. 

Our results indicate that current and forthcoming LHCb datasets have realistic potential to confirm the long-sought $\Xi_{cc}^+$, 
thereby completing the ground-state family of doubly charmed baryons and enabling systematic studies of their structure and dynamics.

 \appendix

 \section{
 	Hadron couplings 
 	in heavy quark spin symmetry 
 }\label{HQET_app}
 
Within the heavy-quark spin symmetry, the $D^0$ and $D^{*0 }$ mesons form a doublet, represented by
\begin{equation}
	\begin{aligned}
		H_D &=
		\frac{1+\slashed{v}}{2}
		\left[
		(D^{*0 })^\mu \gamma_\mu
		+ i\gamma_5 D^0
		\right],   \\
		\overline{H}_D &=
		\left[
		(\overline{D}^{*0})^\mu \gamma_ \mu
		+ i\gamma_5 \overline{D}^0
		\right]
		\frac{1+\slashed{v}}{2},
	\end{aligned}
\end{equation}
where $v$ denotes the meson velocity.
Similarly, the $\Xi_{cc}^+$ and its $J=3/2$ partner $\Xi_{cc}^{*+}$ can be combined into
\begin{equation}
	S^\mu
	= \sqrt{\frac{1}{3}}
	(\gamma^\mu + v_i^\mu)
	\gamma_5 \frac{1+\slashed{v}_i}{2}
	\Xi_ {cc}^+
	+ \frac{1+\slashed{v}_ i}{2}
	(\Xi_ {cc}^{* +})^\mu,
\vspace{1pt} 
\end{equation}
where $(\Xi_{cc}^{*+})^\mu$ is the Rarita–Schwinger spinor and $v_i^\mu$ the baryon velocity.
Under $SU(2)_v$ heavy-quark spin symmetry, these fields transform as
\begin{equation}
	H_D \to e^{i\vec{\epsilon}\cdot\vec{S}_v} H_D, \qquad
	S_ \mu \to e^{i\vec{\epsilon}\cdot\vec{S}_ v} S_ \mu,
\end{equation}
with $\vec{S}_ v$ the heavy-quark spin generators.

Treating $\Lambda_c^+$ as a singlet under this symmetry, the effective couplings take the form
\begin{equation}
	\mathcal{L}
	= g_1 
	\overline{\Lambda_c^+}
	\gamma^\mu\gamma_5
	\overline{H}_D
	S_ \mu
	+ g_2 
	\overline{\Lambda_c^+}
	v ^\mu \gamma_5
	\overline{H}_D
	S_ \mu,
\end{equation}
where $g_{1,2}$ are coupling constants. 
In the near on-shell limit $m_{\Xi_{cc}} \simeq m_{\Lambda_c^+} + m_D$, one has $v  = v_i$, and the second term vanishes since $v_i^\mu S_\mu = 0$.
Expanding the result yields
Eq.~\eqref{HQETrelation} numerically.

\section*{Acknowledgements}
This work is supported in part by the National Key Research and Development Program of China under Grant No. 2020YFC2201501 and  the National Natural Science Foundation of China (NSFC) under Grant No. 12347103 and 12205063.


\begin{thebibliography}{99}


\bibitem{Yu:2017zst}
F.~S.~Yu, H.~Y.~Jiang, R.~H.~Li, C.~D.~L{\"u}, W.~Wang and Z.~X.~Zhao,
Chin. Phys. C \textbf{42},  051001 (2018)
[arXiv:1703.09086 [hep-ph]].



\bibitem{LHCb:2017iph}
R.~Aaij \textit{et al.} [LHCb],
Phys. Rev. Lett. \textbf{119},  112001 (2017)
[arXiv:1707.01621 [hep-ex]].


\bibitem{LHCb:2019epo}
R.~Aaij \textit{et al.} [LHCb],
JHEP \textbf{02}, 049 (2020)
[arXiv:1911.08594 [hep-ex]].



\bibitem{LHCb:2018zpl}
R.~Aaij \textit{et al.} [LHCb],
Phys. Rev. Lett. \textbf{121},   052002 (2018)
[arXiv:1806.02744 [hep-ex]].



\bibitem{SELEX:2002wqn}
M.~Mattson \textit{et al.} [SELEX],
Phys. Rev. Lett. \textbf{89}, 112001 (2002)
[arXiv:hep-ex/0208014 [hep-ex]];
A.~Ocherashvili \textit{et al.} [SELEX],
Phys. Lett. B \textbf{628}, 18 (2005)
[arXiv:hep-ex/0406033 [hep-ex]].
 


\bibitem{BaBar:2006bab}
B.~Aubert \textit{et al.} [BaBar],
Phys. Rev. D \textbf{74}, 011103 (2006)
[arXiv:hep-ex/0605075 [hep-ex]].

\bibitem{Belle:2013htj}
Y.~Kato \textit{et al.} [Belle],
Phys. Rev. D \textbf{89},   052003 (2014)
[arXiv:1312.1026 [hep-ex]].

\bibitem{LHCb:2013hvt}
R.~Aaij \textit{et al.} [LHCb],
JHEP \textbf{12}, 090 (2013)
[arXiv:1310.2538 [hep-ex]].




\bibitem{Detmold:2015aaa}
W.~Detmold, C.~Lehner and S.~Meinel,
Phys. Rev. D \textbf{92},   034503 (2015) 
[arXiv:1503.01421 [hep-lat]].

\bibitem{Wang:2017mqp}
W.~Wang, F.~S.~Yu and Z.~X.~Zhao,
Eur. Phys. J. C \textbf{77},   781 (2017)
[arXiv:1707.02834 [hep-ph]].

\bibitem{Cheng:2018mwu}
H.~Y.~Cheng and Y.~L.~Shi,
Phys. Rev. D \textbf{98},   113005 (2018)
[arXiv:1809.08102 [hep-ph]].
 

\bibitem{Cheng:2020wmk}
H.~Y.~Cheng, G.~Meng, F.~Xu and J.~Zou,
Phys. Rev. D \textbf{101},  034034 (2020)
[arXiv:2001.04553 [hep-ph]].

\bibitem{Tong:2021raz}
H.~Z.~Tong and H.~S.~Li,
Commun. Theor. Phys. \textbf{74},  085201 (2022)
[arXiv:2110.01380 [hep-ph]].

\bibitem{Zeng:2022egh}
S.~Zeng, F.~Xu, P.~Y.~Niu and H.~Y.~Cheng,
Phys. Rev. D \textbf{107},   034009 (2023)
[arXiv:2212.12983 [hep-ph]].

\bibitem{Geng:2022uyy}
C.~Q.~Geng, C.~W.~Liu, A.~Zhou and X.~Yu,
Phys. Rev. D \textbf{107},   053008 (2023)
[arXiv:2211.04372 [hep-ph]].


\bibitem{Geng:2022xfz}
C.~Q.~Geng, X.~N.~Jin, C.~W.~Liu, X.~Yu and A.~W.~Zhou,
Phys. Lett. B \textbf{839}, 137831 (2023)
[arXiv:2212.02971 [hep-ph]].


\bibitem{Liu:2023lsg}
H.~Liu, Y.~H.~Zou, Y.~R.~Liu and S.~Z.~Jiang,
Phys. Rev. D \textbf{108},   014032 (2023)
[arXiv:2304.04575 [hep-ph]].



\bibitem{Liang:2023scp}
Z.~R.~Liang, P.~C.~Qiu and D.~L.~Yao,
JHEP \textbf{07}, 124 (2023)
[arXiv:2303.03370 [hep-ph]].




\bibitem{Kiselev:1998sy}
V.~V.~Kiselev, A.~K.~Likhoded and A.~I.~Onishchenko,
Phys. Rev. D \textbf{60}, 014007 (1999)
[arXiv:hep-ph/9807354 [hep-ph]].

\bibitem{Likhoded:1999yv}
A.~K.~Likhoded and A.~I.~Onishchenko,
[arXiv:hep-ph/9912425 [hep-ph]].

\bibitem{Chang:2007xa}
C.~H.~Chang, T.~Li, X.~Q.~Li and Y.~M.~Wang,
Commun. Theor. Phys. \textbf{49}, 993 (2008)
[arXiv:0704.0016 [hep-ph]].

 
\bibitem{Dulibic:2023jeu}
L.~Dulibi{\'c}, J.~Gratrex, B.~Meli{\'c} and I.~Ni{\v{s}}and{\v{z}}i{\'c},
JHEP \textbf{07}, 061 (2023)
[arXiv:2305.02243 [hep-ph]].


\bibitem{Xing:2023kjk}
Y.~Xing and J.~Xu,
Chin. Phys. C \textbf{49},  093109 (2023)
[arXiv:2311.12346 [hep-ph]].


\bibitem{Fajfer:2003ag}
S.~Fajfer, A.~Prapotnik, P.~Singer and J.~Zupan,
Phys. Rev. D \textbf{68}, 094012 (2003)
[arXiv:hep-ph/0308100 [hep-ph]].



\bibitem{Jia:2024pyb}
C.~P.~Jia, H.~Y.~Jiang, J.~P.~Wang and F.~S.~Yu,
JHEP \textbf{11}, 072 (2024)
[arXiv:2408.14959 [hep-ph]].

\bibitem{Geng:2024uxp}
C.~Q.~Geng, X.~N.~Jin, C.~W.~Liu and X.~Yu,
Phys. Rev. D \textbf{110},   113008 (2024)
[arXiv:2409.11374 [hep-ph]].


\bibitem{Hu:2024uia}
X.~H.~Hu, C.~P.~Jia, Y.~Xing and F.~S.~Yu,
Phys. Rev. D \textbf{111},   076002 (2025)
[arXiv:2403.09511 [hep-ph]].



\bibitem{LHCb:2014ofc}
R.~Aaij \textit{et al.} [LHCb],
JHEP \textbf{08}, 143 (2014)
[arXiv:1405.6842 [hep-ex]].





\bibitem{Cheng:2018hwl}
H.~Y.~Cheng, X.~W.~Kang and F.~Xu,
Phys. Rev. D \textbf{97},   074028 (2018)
[arXiv:1801.08625 [hep-ph]].



\bibitem{Liu:2022igi}
C.~W.~Liu and C.~Q.~Geng,
Phys. Rev. D \textbf{107},   013006 (2023)
[arXiv:2211.12960 [hep-ph]].

\bibitem{Hu:2025ajx}
X.~H.~Hu, Q.~Y.~Zhou, Y.~Xing and Y.~J.~Shi,
Eur. Phys. J. C \textbf{85},   625 (2025)
[arXiv:2502.16561 [hep-ph]].



 
\bibitem{ParticleDataGroup:2024cfk}
S.~Navas \textit{et al.} [Particle Data Group],
Phys. Rev. D \textbf{110},  030001 (2024).

 



\bibitem{CMS:2012wje}
S.~Chatrchyan \textit{et al.} [CMS],
Phys. Lett. B \textbf{714}, 136-157 (2012)
[arXiv:1205.0594 [hep-ex]].

\bibitem{Kramer:2018rgb}
G.~Kramer and H.~Spiesberger,
Chin. Phys. C \textbf{42},   083102 (2018)
[arXiv:1803.11103 [hep-ph]].



\end{thebibliography}
\end{document}